# Evolution of the Primary Pulse in 1D Granular Crystals Subject to On-Site Perturbations: Analytical Study

Yuli Starosvetsky
Faculty of Mechanical Engineering,
Technion Israel Institute of Technology,
Technion City, Haifa 32000, Israel

staryuli@tx.technion.ac.il

**Abstract**

Propagation of primary pulse through an un-compressed granular chain subject to external on-site perturbation is studied. Analytical procedure predicting the evolution of the primary pulse is devised for the general form of the on-site perturbation applied on the chain. The validity of the analytical model is confirmed with several specific granular setups such as, chains mounted on the nonlinear elastic foundation, chains perturbed by the dissipative forces as well as randomly perturbed chains. Additional interesting finding made in the present study corresponds to the chains subject to a special type of perturbations including the terms leading to dissipation and those acting as an energy source. It is shown in the study that application of such perturbation may lead to formation of stable stationary shocks acting as attractors for the initially unperturbed, propagating Nesterenko solitary waves. Interestingly enough the developed analytical procedure provides an extremely close estimations for the amplitudes of these stationary shocks as well as predicts zones of their stability. In conclusion we would like to stress that the developed analytical model have demonstrated spectacular correspondence to the results of direct numerical simulations for all the setups considered in the study.

## 1. Introduction

Over the past decade, dynamics of the one dimensional granular crystals has attracted substantial interest from the researchers of quite different scientific areas [1–17] for their unique dynamical properties. As an example of their peculiar dynamical behavior we resort to a one dimensional (1D) case i.e. chain of steel beads excited by the initial impulse. These chains support the formation of an all new type of propagating, highly localized, elastic stress waves. Indeed as it was shown in [1-8] these waves have a final compact support (spanned over 6 beads) independent on their amplitude and constitute a new class of solitary waves sometimes referred to as 'compactons' [18] exhibiting a super-exponential decay in the tails [19].

Recent studies [10-17] of the dynamics of 1D granular crystals have been mainly concerned with the effect of various types of external perturbations as well as structural disorder on the dynamics of granular chains and in particular on the evolution of the propagating, primary pulses exhibiting solitary like behavior. Thus, wave propagation in the tapered ( i.e. granular chains with constantly reducing/increasing masses of granules along the chain ) and decorated granular chains (i.e. granular chains composed of large granules separated by small ones ) have been addressed in [10-11] both analytically and numerically. Analytical approximation developed in these works for the estimation of the maximal pulse velocity recorded on each one of the granules along with its propagation through the tapered, decorated granular chain have demonstrated a good correspondence of the analytic predictions with that of numerical simulations. This approach was based on the binary collision model (BCA), valid in the limit of sufficiently narrow pulses. Another series of experimental, computational and analytical works considered the dynamics of the periodically disordered granular chains (e.g. di-atomic chains, granular containers, etc.) under various conditions of initial pre-compression [12-17]. Dynamics of primary shock waves in the non-compressed granular chain perturbed by a weak dissipation has been considered in [18-20]. Results of these works shed light on the evolution of the primary pulses in the dissipative, 1D granular media and provided some qualitative theoretical estimations for modeling the dissipation in the chain as well as depicting the rate of decay of the primary pulse.

These recent studies have motivated us to devise a generalized analytical approach predicting the evolution of the amplitude of primary pulse as it propagates through the granular chains subject to on-site perturbation of general type. Analytical approach carried out in the study highly differs from the previously developed methods evaluating spatio-temporal evolution of solitary like pulses in the perturbed media. Many of the earlier developed techniques were based on the method of Witham [23] where effective lagrangian is constructed, out of which a set of slowly evolved parameters (e.g. amplitude, wave number and frequency etc.) of nonlinear wave is derived [24, 25]. There exist many different variations of these techniques depicting the evolution of solitary waves in the systems subject to deterministic perturbation [26,27] as well as randomly perturbed ones [28,30]. In fact these methods have led to many breakthroughs in the various fields of modern applied physics such as non-linear optics, solid state physics, physics of polymers, soft matter and more. However, all these methods can be hardly applied on the perturbed, uncompressed granular lattices due to the lack of analyticity of the system under investigation, induced by the possible separation between the neighboring elements. Moreover, application of the aforementioned variational approaches using effective Lagrangian constructed for the perturbed granular chain has lead us to highly unsatisfactory results.

Analytical procedure presented herein is first developed for the generic form of the on-site perturbation applied on the chain. Several examples of application of the developed method are introduced right afterwards and are applied on some particular granular setups related to the real physical systems. Among these applications we consider effects of uniform nonlinear elastic foundation, dissipative forces, linear foundation with randomly distributed stiffness coefficients. In addition to these obvious realistic examples we also consider a special type of perturbation incorporating the two opposing effects, namely the effect of dissipation and energy sourcing (e.g. Van-Der-Pol oscillator). The latest brings to the very interesting effect of formation of stationary shock waves propagating with constant amplitudes and exhibiting solitary like behavior. It is also shown in this study that these stationary regimes can lose their stability resulting in a monotonous decay of the propagating primary pulse. It is important to stress that the developed methodology have found to be a very successful not only in depicting the evolution of primary pulse but also in the correct prediction of the formation and stability of stationary shocks. Authors convinced that this methodology can be further extended for the more complex granular setups such as higher dimensional granular models as well as for the general class of non-linear lattices (e.g. Toda lattices, FPU chains, etc. ).

## 2. Model

In the present work we study the dynamics of uncompressed granular chains subject to the weak, on-site external perturbations. System under consideration is governed by the following normalized set of equations of motion,

$$\ddot{x}_i = \{x_{i-1} - x_i\}^n \theta(x_{i-1} - x_i) - \{x_i - x_{i+1}\}^n \theta(x_i - x_{i+1}) + \varepsilon R_i(x_i, \dot{x}_i) \tag{1}$$

where $x_i$ is the displacement of the i-th element in the chain, $n$ is assumed to be higher than one ($n > 1$), $R_i$ is the on-site perturbation applied on the i[th] element of the chain, $\varepsilon$ is a small system parameter ($0 < \varepsilon << 1$) which controls the magnitude of the external perturbation, $\theta$ is a Heaviside function which accounts for the separation between the neighboring elements in the granular chain.

Dynamical systems falling under the category described by (1) can be found in many practical applications. In fact, in various experimental setups, granular chains are not free but either hanged (e.g. Newton's craddle), mounted on elastic supports or installed on the guided rails. Thus, granular chain installed on linear/nonlinear elastic foundation (e.g. granular chain embedded in elastic matrix, Newton's cradle [31], chain mounted on the flexures etc.) will experience the perturbation in the form: $R_i(x_i)$ which in the case of small deflections ($|x_i| \ll 1$) can be Taylor expanded and expressed by the following polynomial form $R_i(x_i) \cong \alpha x_i + \beta x_i^2 + \gamma x_i^3 + O(x_i^4)$. Moreover, the chains placed in the guided rails or immersed in any viscous substance will always experience the dissipative type of forces coming up from the friction between the granules and the non-smooth surface of the rail, viscous drag forces of the substance, etc. These forces can appear in the very complicated functional forms, however in many practical applications the effects of friction and the drag forces applied on the element moving in the viscous substance are modeled as $R_i(\dot{x}_i) \cong -\alpha \dot{x}_i, \alpha > 0$. There exist many more examples for the granular systems which fit into the form described by (1), some of them will be considered below in Section 4.

The main focus of the current study is in the developing of analytical procedure describing the evolution of the propagating primary pulse once the external perturbation in the form of (1) is applied on the granular chain. It is important to emphasize that none of the existing analytical techniques dealing with the perturbation of the propagating solitary like profiles are applicable for the current system under investigation and this due to the three main reasons; (1) the underlying, unperturbed system ($\varepsilon = 0$) is not integrable and therefore none of the asymptotic methodologies based on the perturbation near the integrable state [27] are applicable (2) absence of pre-compression allows for separation between adjacent elements of the chain resulting in the non-analyticity of the structure (3) The propagating solitary wave solution in the underlying, unperturbed system ( Nesterenko soliton [1,2] ) is strongly localized and spanned over approximately 6 elements of the chain. The second and the third reasons make the alternative variational methods based on Witham theory [23] (where the integarbility of an underlying system is not required) also hardly applicable leading to the strong deviation of the analytical prediction and numerical simulation.

Therefore a somewhat different analytical approach is developed in the following section in order to correctly depict the evolution of the primary pulse as it propagates through the perturbed granular chain.

## 3. Generalized Analytical Approximation

In the present section we develop an analytical procedure to depict the evolution of the amplitude of the propagating primary pulse interacting with various types of structural defects and in-homogeneities given in the form of (1). The proposed analytical approach is based on the assumption that the functional form of the leading edge profile can be approximated by the solitary wave solution (Nesterenko et.al. [1,2]) derived for the unperturbed case ($\varepsilon \to 0$). It is important to note that the analysis presented in the paper has been developed for the systems represented in coordinates of relative displacements rather than the original ones. Coordinate of the relative displacement is defined as follows

$$\delta_i = x_i - x_{i+1} \tag{2}$$

Thus, substitution of (2) into (1) yields,

$$\ddot{\delta}_i = \delta_{i-1}{}^n \theta(\delta_{i-1}) - 2\delta_i{}^n \theta(\delta_i) + \delta_{i+1}{}^n \theta(\delta_{i+1}) + \varepsilon \{ R_i(x_i, \dot{x}_i) - R_{i+1}(x_{i+1}, \dot{x}_{i+1}) \} \tag{3}$$

As it is evident from (3) the general perturbation term ($\varepsilon \{ R_i(x_i, \dot{x}_i) - R_{i+1}(x_{i+1}, \dot{x}_{i+1}) \}$) cannot be represented solely in terms of relative displacements. However, as it will become clear from the analytical procedure presented below, the analysis performed on (3) will allow the introduction of the locally valid approximations expressing the original coordinates ($x_i$) in terms of the relative displacements ($\delta_i$).

As it was already mentioned above, the method developed herein is based on the unperturbed solution of (3) ($\varepsilon \to 0$) which is a well-known solitary wave solution [1,2]. Before proceeding with the analysis of the perturbed System (3) let us briefly re-derive the recently developed, discrete analytical approximation (i.e. no reliance on long wave approximation) depicting the profile of Nesterenko solitary wave solution supported by (1) in the absence of perturbation ($\varepsilon \to 0$) [1,2]. It is quite clear that an assumption of the propagating traveling wave in the unperturbed, discrete System (2) will reduce it to the following compact form of the advance - delay equation,

$$\ddot{\delta} = \delta(t-T)_+^n - 2\delta(t)_+^n + \delta(t+T)_+^n \tag{4}$$

where for the sake of brevity the Heaviside functions were replaced with (+) subscripts. Without loss of generality the further analysis carried out in the paper assumes $n = 3/2$ and this to admit a Hertzian contact law interaction. From the above discussion it is apparent that System (4) supports the propagation of strongly localized, Nesterenko solitary wave. As it was shown in [32] the solitary wave solution of (4) normalized with respect to the phase shift equated to unity ($T = 1$) can be approximated using the trivial Pade' approximant, resulting in the following relatively simple form.

$$\tilde{S}(\tau) \cong \left( \frac{1}{q_0 + q_2 \tau^2 + q_4 \tau^4 + q_6 \tau^6 + q_8 \tau^8} \right)^2 \tag{5}$$

where the constants $q_0, q_2, q_4, q_6, q_8$ are brought in [32]. At this point it is important to note the obvious limitations of the Pade approximation hidden in the fact that the true solitary wave solution of (4) exhibits a super-exponential decay in the tails unlike the functional form brought in (5) which predicts exponential decay. However, as for the approximation of the entire wave front the correspondence between the proposed approximation and the true solution is very good and the error is bounded by the exponentially small deviations in the tails which by no means can be an obstacle for any asymptotical approximation based on (5) as an unperturbed solution.

We note that System (4) is homogeneous and therefore fully re-scalable. Thus, the approximation (5) constitutes the universal approximation for the arbitrary solitary wave scaled with respect to its amplitude. The relation between the amplitude of the solitary wave and its phase shift are given by:

$$T = \tilde{S}(0)^{1/4} A^{-1/4} \tag{6}$$

Therefore the general solitary wave solution propagating through $n^{th}$ bead can be written as:

$$S_n(t) = A\tilde{S}\left(A^{1/4}t - n + \Psi_0\right) \tag{7}$$

where $A$ is an amplitude of the solitary wave, $n$ is an index of the contact between the $n^{th}$ and $(n+1)^{th}$ elements of the granular chain, $\Psi_0$ is the initial phase which can be equated to zero without loss of generality. In Figure 1 we plot the time series of the response recorded on the three adjacent contacts, namely $\delta_{n-1}, \delta_n, \delta_{n+1}$ corresponding to the propagating solitary wave solution in the unperturbed chain ($\varepsilon = 0$) described by (3).

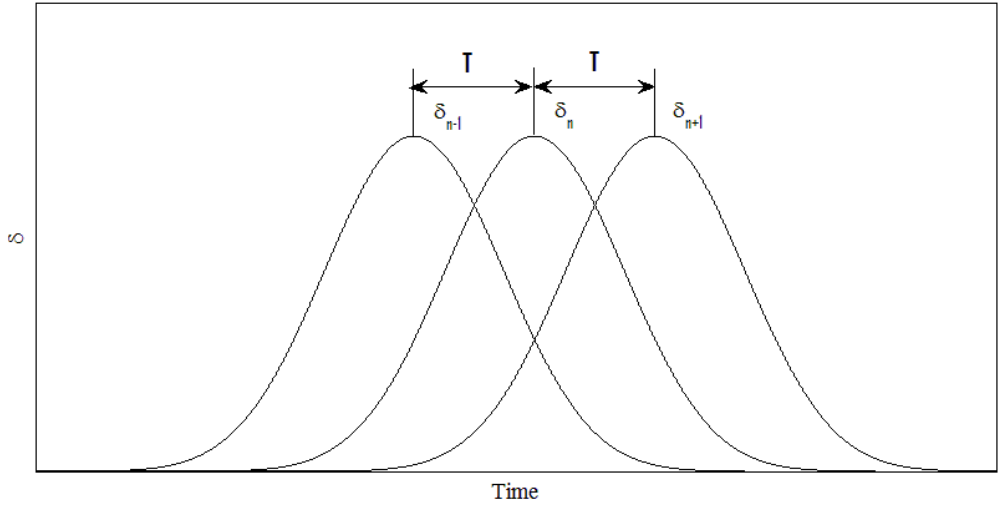

Figure 1 Relative displacement response on the three adjacent contacts

As it was already stressed in the previous section the primary focus of the current work is in devising analytical approach to study the dynamics of the leading edge profile along with its propagation through the perturbed granular chain ($\varepsilon \neq 0$). To visualize that, let us examine the response recorded on the three adjacent contacts of the perturbed granular chain. As a perturbation we choose a uniform linear stratification of the granular chain ($R_i = u_i$) described by the following set of equations,

$$\ddot{\delta}_i = \{\delta^n_{i-1}\}_+ - 2\{\delta^n_i\}_+ + \{\delta^n_{i+1}\}_+ - \varepsilon\{\delta_i\} \tag{8}$$

In Figure 2 we illustrate the response on the three adjacent contacts. The evolution of the leading edge as it propagates from contact to contact which can be clearly seen from the results of Figure 2.

It is quite reasonable to assume the near solitary state behavior on the leading edge of the propagating disturbance. Moreover, as it comes from the results of Figure 2 the residual fluctuations left behind the leading edge are not negligible. Therefore the near solitary like behavior cannot be assumed over the entire space as it is usually done for similar type of problems where the slow evolution of solitary like solution over the slightly perturbed media is studied. Bearing this in mind we develop a local approach for spotting the evolution of the leading edge profile as it propagates through the perturbed, 1D granular media.

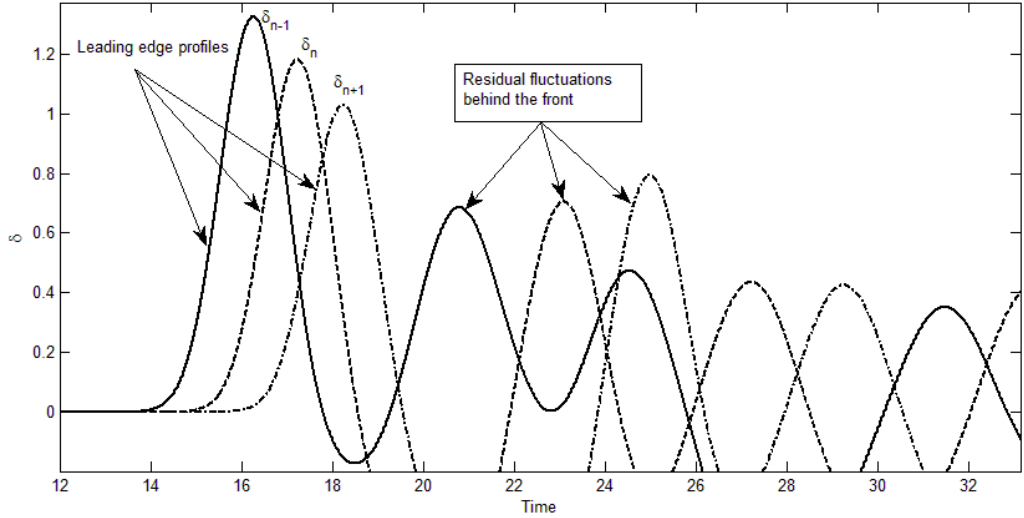

Figure 2 Evolution of the leading edge due to stratification of the granular chain

Let us start the description of the analytical procedure from several definitions and assumptions:

1. For each contact in the chain we assign the local, semi-infinite time interval $(-\infty, T_{sn}]$, such that $T_{sn}$ corresponds to the exact time point when the $n^{th}$ contact reaches the peak on the leading edge (primary pulse) of the propagating disturbance (cf. Figure 4).

2. We assume that in the considered time interval $t \in (-\infty, T_{sn}]$ the response of the $n^{th}$ contact can be approximated as $\delta_n(t) = A_n \tilde{S}\left(A_n^{1/4} t - n\right) + o(\varepsilon)$, $T_{sn} = n A_n^{-1/4}$, the response of the (n-1)$^{th}$ contact can be approximated as $\delta_{n-1}(t) = A_{n-1} \tilde{S}\left(A_{n-1}^{1/4} t - n + 1\right) + o(\varepsilon)$ and that of (n+1)$^{th}$ contact as $\delta_{n+1}(t) = A_{n+1} \tilde{S}\left(A_{n+1}^{1/4} t - n - 1\right) + o(\varepsilon)$

3. For the time interval $t \in (-\infty, T_{sn}]$ we assume that the response of the (n+2)$^{th}$ contact ($\delta_{n+2}(t)$) and its successors ($\delta_{n+3}(t), \delta_{n+4}(t), \ldots$) as well as that of (n+2)$^{th}$ granule ($u_{n+2}(t)$) and the rest of the elements of higher index ($u_{n+3}(t), u_{n+4}(t), \ldots$) are negligibly small $o(\varepsilon)$.

In fact assumption (3) is a bit tricky and deserves some extra explanations. If we look on the response on the 4 adjacent contacts of the unperturbed System illustrated in Figure 3 we observe that during the time interval $t \in (-\infty, T_{si}]$ the response recorded on the (i+2)$^{th}$ contact is negligibly small $o(\varepsilon)$ and this due to the super-exponential decay exhibited by the compactons in the tails [22] (in fact it is an exponential decay in case of the proposed approximation (5) used instead of the true solution). However, the magnitude of the response assumed to be a transcendentally small terms which by no means can affect an error of the proposed analytical approximation.

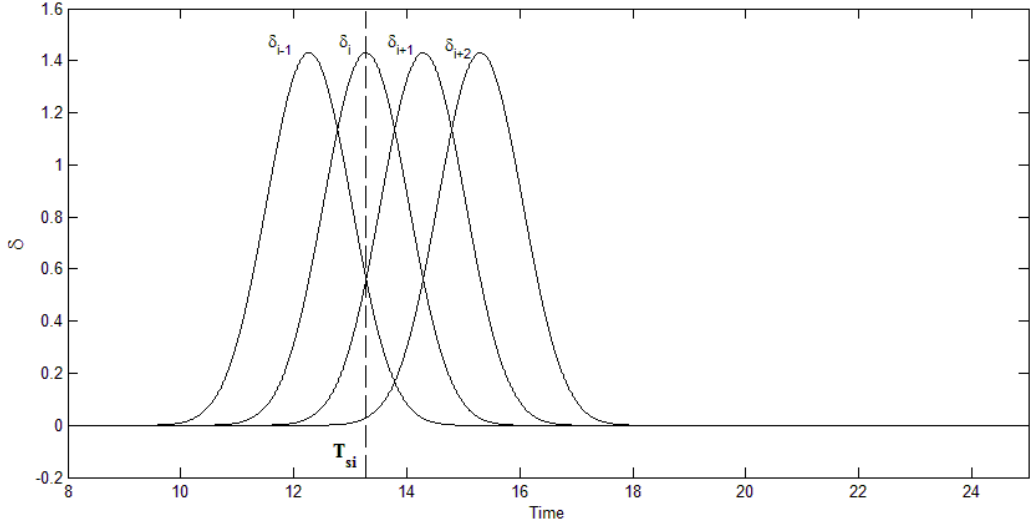

Figure 3 Response on the 4 successive contact elements

4. Clearly enough the original coordinate $u_i$ (displacement of the i$^{th}$ bead) can be expressed as: $u_i = (u_i - u_{i+1}) + (u_{i+1} - u_{i+2}) + (u_{i+2} - u_{i+3}) + u_{i+3} = \delta_i + \delta_{i+1} + \delta_{i+2} + u_{i+3}$. However, as it was assumed in (3) for the considered time interval $t \in (-\infty, T_{sn}]$ the response of $u_{i+2}, u_{i+3}, \ldots$ as well as that of $\delta_{i+2}, \delta_{i+3}, \ldots$ is negligibly small (exponentially small) and therefore can be omitted from the analysis for the considered time interval. Thus, the original displacement of the i$^{th}$ bead can be approximated in terms of relative displacements:

$$u_i = \delta_i + \delta_{i+1} + o(\varepsilon) \qquad (9)$$

where $\delta_{i+2}(t)$ is assumed to be of $o(\varepsilon)$ for $t \in (-\infty, T_{sn}]$.

5. The last assumption adopted for the analysis is that the perturbation term $R_i(u_i, \dot{u}_i)$ introduced in (1) is assumed to be of an arbitrary polynomial form i.e.

$$R_i(u_i, \dot{u}_i) = G_0^i + G_1^i u_i + G_2^i \dot{u}_i + G_3^i u_i^2 + G_4^i u_i \dot{u}_i + G_5^i \dot{u}_i^2 + \ldots \qquad (10)$$

Thus, substituting (9) into (3) yields:

$$\ddot{\delta}_i = \{\delta_{i-1}{}^n\}_+ - 2\{\delta_i{}^n\}_+ + \{\delta_{i+1}{}^n\}_+ + \varepsilon\{R_i(\delta_i + \delta_{i+1}, \dot{\delta}_i + \dot{\delta}_{i+1}) - R_{i+1}(\delta_{i+1} + \delta_{i+2}, \dot{\delta}_{i+1} + \dot{\delta}_{i+2})\} + o(\varepsilon),$$
$$t \in [-\infty, T_{si}]$$

(11)

It is important to note that Eq. (11) is valid only in the time interval $t \in (-\infty, T_{si}]$, namely for the characteristic time period of the primary shock, propagating through the i$^{th}$ contact. Thus accounting for the assumptions described in (3), Eq. (11) can be simplified even further by equating all the terms containing $\delta_{i+2}$ to zero.

$$\ddot{\delta}_i = \delta_{i-1}{}^n \theta(\delta_{i-1}) - 2\delta_i{}^n \theta(\delta_i) + \delta_{i+1}{}^n \theta(\delta_{i+1}) + \varepsilon\{R_i(\delta_i + \delta_{i+1}, \dot{\delta}_i + \dot{\delta}_{i+1}) - R_{i+1}(\delta_{i+1}, \dot{\delta}_{i+1})\} + o(\varepsilon),$$
$$t \in [-\infty, T_{si}]$$

(12)

It is important to emphasize that by omitting the terms containing $\delta_{i+2}$ we assume that the order of magnitude of $\delta_{i+2}$ is lower than that prescribed for the perturbation $(\delta_{i+2} \sim o(\varepsilon))$ which is quite a reasonable assumption for the case of super-exponentially decaying tails exhibited by compactons [21, 22].

Accounting for all the definitions and assumptions listed above (1-5) we are in the point to formulate the analytical procedure to estimate the evolution of the amplitude of the primary pulse along with its propagation through the perturbed granular chain.

To this end we start from the assumption (2) that the leading edge profile can be approximated by the solitary wave solution (11) in the time interval $(-\infty, T_{sn}]$ (See Figure 4).

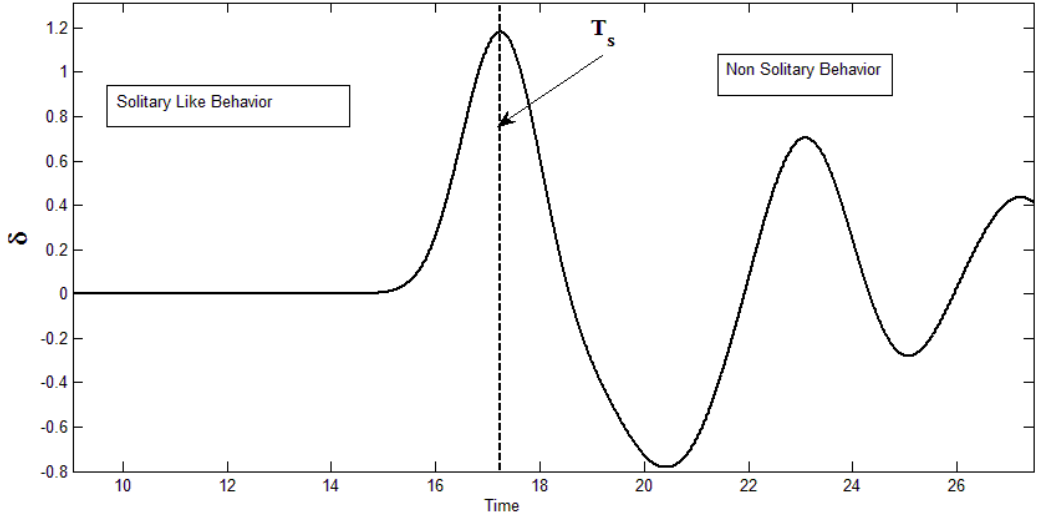

Figure 4 Response of the 15$^{th}$ contact of System (8).

Let us write down again the perturbed System

$$\ddot{\delta}_i = \delta_{i-1}^n - 2\delta_i^n + \delta_{i+1}^n + \varepsilon g_i\left(\delta_i, \delta_{i+1}, \dot{\delta}_i, \dot{\delta}_{i+1}\right) + o(\varepsilon) \qquad (13)$$

where $g_i = \{R_i(\delta_i + \delta_{i+1}, \dot{\delta}_i + \dot{\delta}_{i+1}) - R_{i+1}(\delta_{i+1}, \dot{\delta}_{i+1})\}$

As it was already mentioned above the solution of (13) is assumed to be in the form of a solitary wave:

$$\delta_i(t) = S_i(t) = A_i \tilde{S}\left(A_i^{1/4} t - i\right) \qquad (14)$$

which is valid in the interval $(-\infty, T_{si}]$. As we assume the general waveform of the solution to be known, the parameter that defines the solution on the leading edge is $A_i$ which is declared as an unknown in the developed approximation.

Introducing (14) in (13) yields the following System of equations,

$$A_i^{3/2}\tilde{S}_i'' = A_{i-1}^{3/2}\tilde{S}_{i-1}^{3/2} - 2A_i^{3/2}\tilde{S}_i^{3/2} + A_{i+1}^{3/2}\tilde{S}_{i+1}^{3/2} + \varepsilon g_i(A_i\tilde{S}_i, A_{i+1}\tilde{S}_{i+1}, A_i^{5/4}\tilde{S}_i', A_{i+1}^{5/4}\tilde{S}_{i+1}')$$
$$t \in (-\infty, T_{si}] \qquad (15)$$

As it was already mentioned above Equation (15) is an approximation for the solitary like behavior on the leading edge of the response which is valid only for the time interval $(-\infty, T_{si}]$. Our main purpose in the current approach is a determination of the set of unknowns $\{A_i\}$ which defines the evolution of the primary pulse along with its propagation through the perturbed granular chain. It is quite clear that a straightforward determination of the required set of parameters is impossible if all the contacts are taken into consideration simultaneously. However, recalling that the current approximation is developed for the strongly localized pulses (compactons) and for the specific time interval $(-\infty, T_{si}]$ we construct an iterative procedure, which estimates in each step the amplitude $A_i$ of the leading edge propagating through the i$^{th}$ contact.

The next step of the approximation is a direct integration of (15) in the time interval $(-\infty, T_{si}]$ leading to the following expression,

$$0 = A_{i-1}^{3/2} \int_{-\infty}^{T_{si}} \tilde{S}_{i-1}^{3/2} dt - 2A_i^{3/2} \int_{-\infty}^{T_{si}} \tilde{S}_i^{3/2} dt + A_{i+1}^{3/2} \int_{-\infty}^{T_{si}} \tilde{S}_{i+1}^{3/2} dt + \varepsilon \int_{-\infty}^{T_{si}} g_i(A_i \tilde{S}_i, A_{i+1} \tilde{S}_{i+1}, A_i^{5/4} \tilde{S}_i', A_{i+1}^{5/4} \tilde{S}_{i+1}') dt \quad (16)$$

It is important to note that the inertia term in (15) vanishes right after the integration in the specified time interval. This is clear from the fact that $T_{Si}$ is the time chosen on the peak of the leading edge where the first derivative vanishes $\left(\dot{\delta}_i(t)|_{t=T_{Si}} = 0\right)$.

By a proper scaling and shifting of time in the integrals of the first three terms in (18) one obtains the following simplification,

$$0 = A_{i-1}^{5/4} f_1 - 2A_i^{5/4} f_2 + A_{i+1}^{5/4} f_3 + \varepsilon \int_{-\infty}^{T_{si}} g_i(A_i \tilde{S}_i, A_{i+1} \tilde{S}_{i+1}, A_i^{5/4} \tilde{S}_i', A_{i+1}^{5/4} \tilde{S}_{i+1}') dt$$

$$f_1 = \int_{-\infty}^{0} \tilde{S}(t+1)^{3/2} dt, \; f_2 = \int_{-\infty}^{0} \tilde{S}(t)^{3/2} dt, \; f_3 = \int_{-\infty}^{0} \tilde{S}(t-1)^{3/2} dt$$

(17)

where, $f_1, f_2, f_3$ are universal constants. It is also important to stress at this point that if the perturbation $R_i$ is chosen to be a linear function ($R_i(u_i, \dot{u}_i) = g_0 + g_1 u_i + g_2 \dot{u}_i$), the same simplification can be applied also on the forth term providing a simple two dimensional map. However, solving (17) for the general nonlinear case is quite a formidable task. Therefore, to proceed with the analysis of (17) we require additional asymptotical assumption. For that matter it would be quite reasonable to assume that the variation of the amplitude of the leading edge profile propagating from contact to contact is bounded by the order of applied perturbation, namely $O(\varepsilon)$. Therefore, the amplitude $A_{i+1}$ can be approximated as,

$$A_{i+1} = A_i + \varepsilon \Delta + o(\varepsilon) \qquad (18)$$

Thus accounting for (18) in (17) yields,

$$0 = A_{i-1}^{5/4} f_1 - 2 A_i^{5/4} f_2 + A_{i+1}^{5/4} f_3 + \varepsilon \int_{-\infty}^{T_{si}} g_i \left( A_i \tilde{S}_i, A_i \tilde{S}_{i+1}, A_i^{5/4} \tilde{S}_i{}', A_i^{5/4} \tilde{S}_{i+1}{}' \right) + o(\varepsilon)$$

$$f_1 = \int_{-\infty}^{0} \tilde{S}(t+1)^{3/2} dt, \; f_2 = \int_{-\infty}^{0} \tilde{S}(t)^{3/2} dt, \; f_3 = \int_{-\infty}^{0} \tilde{S}(t-1)^{3/2} dt \qquad (19)$$

Here we should make a note that the further developed analytical procedure applied on (19) assumes that the value of the amplitude on the (i-1)$^{th}$ contact ($A_{i-1}$) is known. Therefore, $A_i$ and $A_{i+1}$ are to be determined in the subsequent steps. To this extent additional equation is required for the considered time interval $(-\infty, T_{si}]$ to find the pair of unknowns ($A_i$, $A_{i+1}$). This equation is derived straightforwardly by resorting to the equation of motion for the (i+1)$^{th}$ contact <u>with respect to the same time interval $(-\infty, T_{si}]$</u>.

$$\ddot{\delta}_{i+1} = \delta_i{}^n \theta(\delta_i) - 2\delta_{i+1}{}^n \theta(\delta_{i+1}) + \delta_{i+2}{}^n \theta(\delta_{i+2}) + \varepsilon g_{i+1}\left(\delta_{i+1}, \dot{\delta}_{i+1}, \delta_{i+2}, \dot{\delta}_{i+2}\right) + o(\varepsilon)$$

$$t \in [-\infty, T_{si}] \qquad (20)$$

Recalling that the amplitude of the response recorded on the (i+2)$^{th}$ contact is assumed to be of $o(\varepsilon)$ (for the considered time interval $(-\infty, T_{si}]$) then all the terms in (20) containing $\delta_{i+2}$ are omitted. This simplification yields the following equation,

$$\ddot{\delta}_{i+1} = \delta_i{}^n \theta(\delta_i) - 2\delta_{i+1}{}^n \theta(\delta_{i+1}) + \varepsilon g_{i+1}\left(\delta_{i+1}, \dot{\delta}_{i+1}\right) + o(\varepsilon)$$

$$t \in [-\infty, T_{si}] \qquad (21)$$

Thus, applying the same procedure as before on Eq. (25) yields,

$$A_i^{5/4}(-f_4 + f_5) - 2 A_{i+1}^{5/4} f_6 + \varepsilon \int_{-\infty}^{T_{si}} g_{i+1}\left(A_{i+1}\tilde{S}_{i+1}, A_{i+1}^{5/4}\tilde{S}_{i+1}{}'\right) dt + o(\varepsilon) = 0$$

$$f_4 = \int_{-\infty}^{0} \tilde{S}(t-1)'' dt, \; f_5 = \int_{-\infty}^{0} \tilde{S}(t)^{3/2} dt, \; f_6 = \int_{-\infty}^{0} \tilde{S}(t-1)^{3/2} dt, \qquad (22)$$

Thus, at each step of the iterative process the combined system (19) and (22) is solved and this to obtain the required amplitudes of the primary pulse $\{A_i\}$.

Recalling that the perturbation is assumed to have a polynomial form, Eq. (23) can be rewritten as follows,

$$A_{i-1}^{5/4} f_1 - 2 A_i^{5/4} f_2 + A_{i+1}^{5/4} f_3 + \varepsilon A_i^{k+l+5/4(m+n)-1/4} \sum_{k,l,m,n} G_{klmn}^{(1)i} g_{klmn}^{(1)} = 0$$

$$A_i^{5/4}(f_5 - f_4) - 2 A_{i+1}^{5/4} f_6 + \varepsilon A_{i+1}^{p+5/4r-1/4} \sum_{p,r} G_{pr}^{(2)i} g_{pr}^{(2)} = 0 \quad (23)$$

$$g_{klmn}^{(1)} = \int_{-\infty}^{0} S(t)^k S(t-1)^l S(t)'^m S(t-1)'^n dt, \quad g_{pr}^{(2)} = \int_{-\infty}^{0} S(t-1)^p S(t-1)'^r dt$$

Where $\{G_{klmn}^{(1)i}\}, \{G_{pr}^{(2)i}\}$ are the sets of arbitrary constants defined by the perturbation. Though System (23) constitutes a two-dimensional map we use it solely for estimating the amplitude of the primary pulse on the i$^{th}$ contact $A_i$. The value of $A_{i+1}$ obtained from (23) in the same step of the iterative procedure is not of use and is found only in the next iteration of the map. The reason for that is simple. In fact for the considered time interval $(-\infty, T_{si}]$ the response of the (i+1)$^{th}$ contact only starts building up and due to the compacton like behavior away from the peak [22] is highly insensitive to the value of the amplitude on the peak ($A_{i+1}$). Therefore, reliance on the value of $A_{i+1}$ calculated on the (i)$^{th}$ step of the map may introduce a considerable error to the scheme. In fact $A_{i+1}$ is a dummy variable which is used in System (23) for estimation of $A_i$. Moreover, pursuing this study we have made a very important and interesting observation suggesting that $A_{i+1}$ can be replaced with $A_i$ in the first equation of (23) itself causing almost no change in the performance of the map and this due to the strong insensitivity of the response in initial phase (region of a super-exponential decay) on the (i+1)$^{th}$ contact to the variation of its amplitude at the peak of the leading edge. However, in the course of the present paper we use the two dimensional map everywhere as it can be better justified asymptotically.

The proposed analytical procedure evaluating the set of amplitudes $\{A_i\}$ of the primary pulse can be summarized in the three main stages:

(1) Initialization of the scheme (27) by assigning $A_{i-1}$ the initial value of the amplitude of the primary pulse (at the initial stage of the process)

(2) Solution of (27) and derivation of the pair ($A_i, A_{i+1}$)

(3) Assigning $A_{i-1} \leftarrow A_i$ and moving back to the second stage (2)

In the next section we demonstrate that for some particular types of perturbation (uniform perturbation), scheme (23) can be approximated analytically - providing spectacular correspondence between the numerical simulation of the true system, recursive approximation obtained from the map (23) as well as an explicit analytical expression depicting the evolution of the amplitude of the primary pulse.

Let us close the current section by stressing that the developed analytical approximation is by no means restricted to the Hertzian type of interaction and can be applied for any type of nonlinear interacting potential ($n \geq 5/2$) between the neighboring elements of granular chain. The only reason for assuming Hertzian interaction law in this study is that it can be easily realized in various practical applications.

## 4. Applications

In the present section we apply the proposed analytical procedure on systems comprising granular chains subject to various types of on-site perturbations. In all the configurations considered herein we apply the initial impulse on the first element of the granular chain and depict its evolution in space along with its interaction with the perturbation. Moreover, the initial part of the chain is left unperturbed when after a certain element index the perturbation is activated (cf. Figure 5). This is done to assure the formation of a pure (unperturbed) Nesterenko soliton before entering the perturbed zone of the chain.

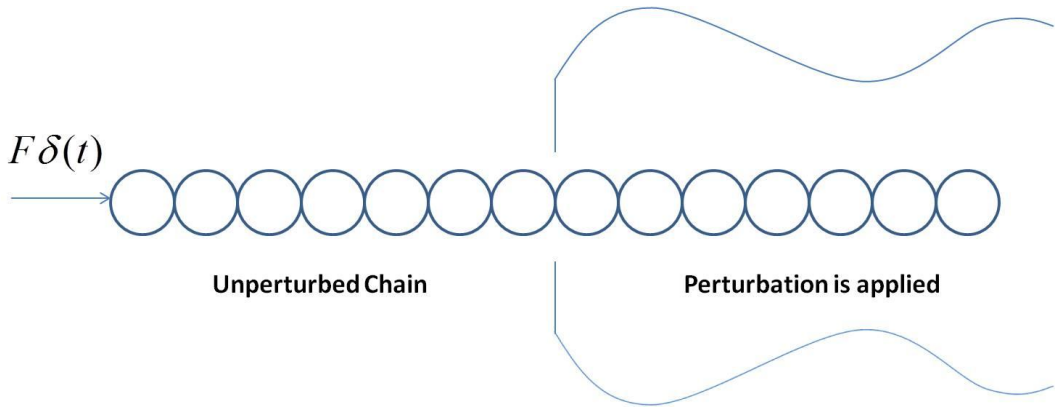

Figure 5 General scheme of the perturbed granular chain

Various types of applied perturbation rising in the real granular setups are considered. Among those the effects of deterministic and random elastic foundation (stratified potentials), friction and drag forces on the evolution of the primary pulse are studied.

## 4.1 Effect of the Uniform Linear/Nonlinear Elastic Foundation on the Primary Pulse Propagation

Let us start from the application of the proposed methodology on the very basic dynamical system comprising the regular, uncompressed, one-dimensional granular chain mounted on linear/non-linear elastic foundation subject to initial impulse (Fig. 5). Equations of motion for the system under consideration are as following,

$$\ddot{x}_i = \{x_{i-1} - x_i\}^n \theta(x_{i-1} - x_i) - \{x_i - x_{i+1}\}^n \theta(x_i - x_{i+1}) + \varepsilon\{\alpha x_i + \beta x_i^3\} \tag{24}$$

In fact for the sake of demonstration we choose the effect of elastic foundation to be modeled as a cubic polynomial; however it can assume any arbitrary polynomial forms. Again we are interested in depicting the propagation of the primary pulse through the contacts of granules and henceforth rewrite (24) in terms of relative displacement. Generally the description of the dynamics of shock waves in terms of relative displacements is of practical significance as it is directly related to the measure of the stress wave propagating through the chain. Arguing as above System (24) is rewritten as following,

$$\ddot{\delta}_i = \{\delta_{i-1}\}_+^n - 2\{\delta_i\}_+^n + \{\delta_{i+1}\}_+^n + \varepsilon\{\alpha \delta_i + \beta(\delta_i^3 + 3\delta_i^2 \delta_{i+1} + 3\delta_i \delta_{i+1}^2)\} \tag{25}$$

Application of the procedure of the previous section on System (25) leads to the two-dimensional map,

$$\begin{aligned}
0 &= f_1 A_{i-1}^{5/4} - 2 f_2 A_i^{5/4} + f_3 A_{i+1}^{5/4} + \varepsilon\left(\alpha g_{1000}^{(1)} A_i^{3/4} + \beta\left(g_{3000}^{(1)} + 3 g_{2100}^{(1)} + 3 g_{1200}^{(1)}\right) A_i^{11/4}\right) \\
0 &= (f_5 - f_4) A_i^{5/4} - 2 f_6 A_{i+1}^{5/4} + \varepsilon\left(\alpha g_{10}^{(2)} A_{i+1}^{3/4} + \beta g_{30}^{(2)} A_{i+1}^{11/4}\right)
\end{aligned} \tag{26}$$

Fortunately for the case of uniformly perturbed chains (i.e. $\alpha$, $\beta$ remain constant all through the chain) the map (26) can be homogenized. The homogenization is applied on the first of the equations of (26) which is primarily related to the dynamics of the i$^{th}$ contact.

Thus performing the transformation $A_i^{5/4} = B_i$ leads to,

$$f_1 B_{i-1} - 2 f_2 B_i + f_3 B_{i+1} + \varepsilon\left(\alpha g_{1000}^{(1)} B_i^{3/5} + \beta\left(g_{3000}^{(1)} + 3 g_{2100}^{(1)} + 3 g_{1200}^{(1)}\right) B_i^{11/5}\right) = 0 \tag{27}$$

Long wave approximation yields the following transformation,

$$\begin{aligned}
B_{i-1} &= B - hB' + O(h^2) \\
B_{i+1} &= B + hB' + O(h^2)
\end{aligned} \tag{28}$$

Substitution of (28) into (27) results in a simple ordinary differential equation of the first order,

$$h(f_1 - f_3)B' = \varepsilon \left( \alpha g^{(1)}_{1000} B^{3/5} + \beta \left( g^{(1)}_{3000} + 3 g^{(1)}_{2100} + 3 g^{(1)}_{1200} \right) B^{11/5} \right) \tag{29}$$

It is obvious that in the normalized system under investigation $h = 1$ and thus System (29) is rewritten as,

$$B' = \varepsilon^* \left( \alpha g^{(1)}_{1000} B^{3/5} + \beta \left( g^{(1)}_{3000} + 3 g^{(1)}_{2100} + 3 g^{(1)}_{1200} \right) B^{11/5} \right)$$
$$\varepsilon^* = \varepsilon / (f_1 - f_3) \tag{30}$$

Separation of variables of (30) leads to the exact solution represented in quadrature form,

$$\int \frac{dB}{\left( \alpha f_{1000} B^{3/5} + \beta \left( f_{3000} + 3 f_{2100} + 3 f_{1200} \right) B^{11/5} \right)} = \varepsilon^* t + C \tag{31a}$$

Apparently, integral appearing in the LHS of (31a) can be calculated exactly; however this leads to a very cumbersome and lengthy expression which is abandoned in the analysis. However, for the two limiting cases, namely for the case of the linear stratification ($\beta = 0$) as well as purely non-linear one ($\alpha = 0$) simple explicit analytical solutions are derived.

Thus, starting with the linear case ($\beta = 0$) the evolution of the amplitude of the propagating primary pulse reads,

$$A = \left\{ A_0^2 + \frac{8}{5} \varepsilon^* \alpha f_{1000} t \right\}^{1/2} \tag{31b}$$

where $A_0$ is the initial amplitude of the primary pulse.

As for the purely nonlinear case ($\alpha = 0$) the amplitude of the propagating primary pulse reads,

$$A = \left\{ A_0^{-6/4} + \frac{6\varepsilon^* \beta}{5} \left( f_{3000} + 3 f_{2100} + 3 f_{1200} \right) t \right\}^{-4/6} \tag{31c}$$

In Figures 6,7 we illustrate the comparison of the analytical prediction for the evolution of the amplitude of the primary pulse obtained from the map (26) and from the long wave approximation (31) with that of a direct numerical simulation of the original system. Results of comparison brought in Figure 6 correspond to the case of linear elastic foundation $\beta = 0, \alpha < 0$. The case of purely nonlinear elastic foundation is illustrated in Figure 7. For both the cases (linear, purely nonlinear elastic foundation) an identical initial impulse is applied on the left part of the chain where the first 16 elements of the chain are kept unperturbed and the perturbation is only activated starting from the 16[th] element. As it was already explained above the initial unperturbed layer is introduced in order to secure the formation of the pure Nesterenko solitary wave [1, 2] - farther evolution of which (due to the applied perturbation) is analyzed.

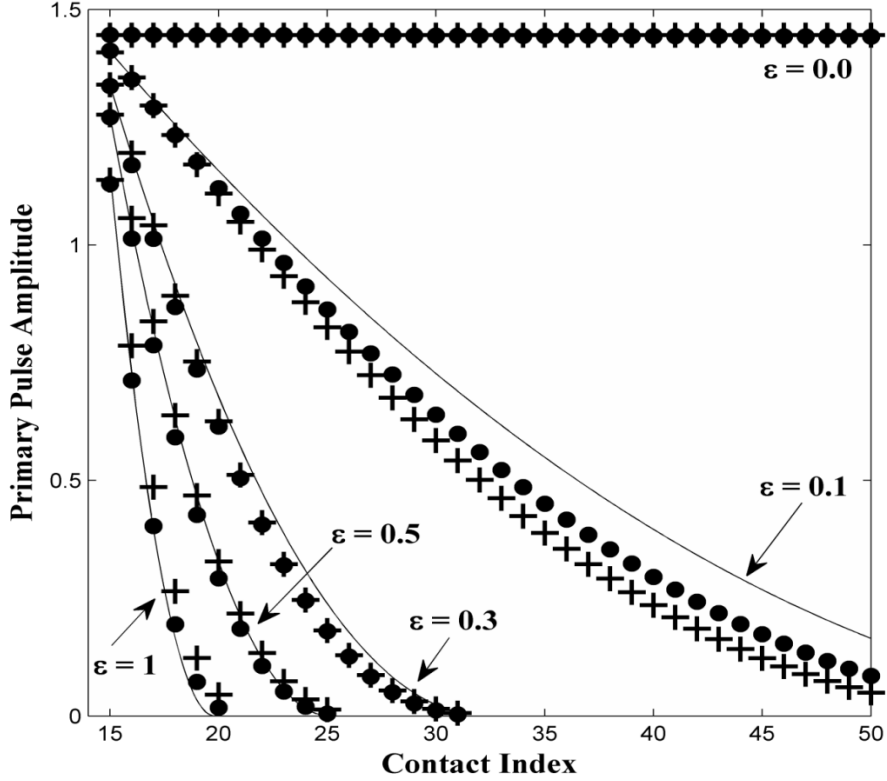

Figure 6 Comparison of analytical prediction and numerical simulation for the various values of $\varepsilon$ parameter ($\varepsilon = 0.1, 0.3, 0.5, 1$), System Parameters: $\beta = 0, \alpha = -1$. Amplitudes of the primary pulse recorded on the contacts from the direct numerical simulation are denoted with (+) signs, those predicted by the map (26) are denoted with the bolded dots (.) and finally the amplitude evolution predicted by the long wave approximation (31b) is denoted with the thin solid line (-).

We note that in order to compare the set of amplitudes $\{A_i\}$ predicted by the map (26) with the peak values of the numerically simulated responses the peak value of the response of the i[th] contact should be calculated from the approximation,

$$\delta_i(t)|_{t=T_{Si}} = A_i \tilde{S}(0) = \frac{A_i}{q_0^2} \tag{32}$$

Where $\delta_i(t)|_{t=T_{Si}}$ is the value of the response in the peak. Therefore, the set of amplitudes $\{A_i\}$ obtained from the map is re-scaled with respect to (36) to provide a fair comparison with the peak of actual response.

Results of Figures 6, 7 suggest for the good agreement between the analytical predictions of the evolution of the primary pulse with that derived from numerical simulation. It is also important to note that for the case of purely nonlinear elastic foundation we didn't include the same comparison tests as for the linear one and thus included only the two main limiting cases; $\varepsilon = 0.1$ (which is situated far away from the asymptotical limit of validity epsilon = 1) and $\varepsilon = 1$ which is exactly the limit. The reason for omitting another two comparison tests (epsilon=0.3, epsilon=0.5) is because of the densification of the data in the narrow region. This may leave a room for confusion in the interpretation of the results.

Surprisingly enough the proposed analytical approximation is in a good agreement with the results of numerical simulation even in the limit of validity of the fundamental asymptotical assumption (epsilon<<1), namely for the case of epsilon = 1 (Figs. 6, 7). This agreement holds for both linear and purely non-linear cases.

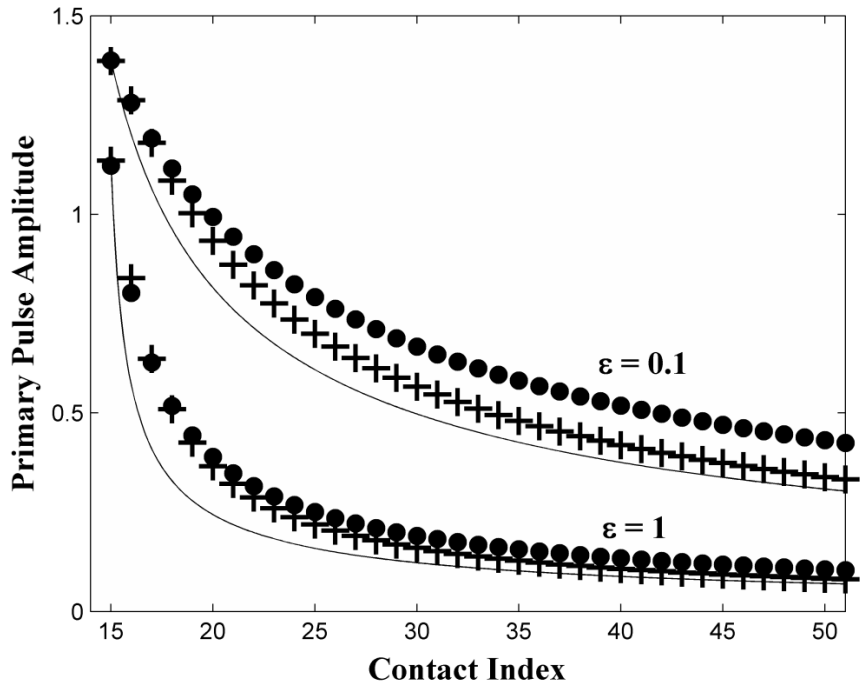

Figure 7 Comparison of analytical prediction and numerical simulation for the various values of $\varepsilon$ parameter ($\varepsilon = 0.1, 0.3, 1.0$) $\beta = -1, \alpha = 0$. Amplitudes of the primary pulse recorded on the contacts of the chain from the direct numerical simulation are denoted with (+) signs, those predicted by the map

(26) are denoted with the bolded dots (.) and finally the amplitude evolution predicted by the long wave approximation (31c) is denoted with the thin solid line (-). Note that the set of amplitudes $\{A_i\}$ obtained from the map (26) is re-scaled with respect to (32) to provide a fair comparison with the peak of actual response.

Summarizing the results of the present subsection we would like to emphasize once again that the analytical procedure illustrated for the two limiting cases namely for the case of linear and purely nonlinear, uniform elastic foundations is general and can be applied on the arbitrary polynomial form. In fact it is of common practice in the real physical systems to model the effects of elastic foundation by a certain polynomial form under assumption of small deflections. This makes the analysis developed above of broad applicability for the homogeneous, strongly nonlinear lattices perturbed by any type of linear/ non-liner elastic foundation.

4.2 Effect of a Dissipative Force on the Primary Pulse Propagation

In the present subsection we apply the developed analytical methodology to depict the evolution of a primary pulse propagating through the granular chain subject to the viscous dissipative force, which may rise due to the interaction of the chain with the surrounding viscous substance. The system under consideration is described by the following set of equations,

$$\ddot{x}_i = \{x_{i-1} - x_i\}^n \theta(x_{i-1} - x_i) - \{x_i - x_{i+1}\}^n \theta(x_i - x_{i+1}) - \varepsilon \mu \dot{x}_i \quad (33)$$

As it is already apparent from the analysis brought above, we first need to rewrite (33) in terms of the relative displacements leading to the following set of equations of motion,

$$\ddot{\delta}_i = \{\delta_{i-1}\}_+^n - 2\{\delta_i\}_+^n + \{\delta_{i+1}\}_+^n - \varepsilon \mu \dot{\delta}_i \quad (34)$$

Applying the analytical procedure as before, System (34) is reduced to the following map,

$$\begin{aligned} 0 &= f_1 A_{i-1}^{5/4} - 2 f_2 A_i^{5/4} + f_3 A_{i+1}^{5/4} + \varepsilon \alpha\, g^{(1)}_{0010} A_i \\ 0 &= (f_5 - f_4) A_i^{5/4} - 2 f_6 A_{i+1}^{5/4} + \varepsilon \alpha\, g^{(2)}_{01} A_{i+1} \end{aligned} \quad (35)$$

Homogenization of the map (35) is performed exactly as above leading to the following explicit analytical expression depicting the evolution of the primary pulse due to the linear dissipation acting on the chain,

$$A = \left\{ A_0^{1/4} + \frac{\varepsilon^* \alpha\, g^{(1)}_{0010}}{5} t \right\}^4 \quad (36)$$

In Figure 8 we illustrate the comparison between the amplitude evolution of the primary pulse predicted by the map (35) as well as by the continuum limit approximation (36) with that derived from the direct numerical simulation of (33). It is important to note that the numerical procedure applied on System (33) was performed similarly to the one described in the previous sub-section. Thus, initial impulse is applied on the left part of the chain where the first 16 elements of the chain are kept unperturbed and the perturbation is only activated starting from the $16^{th}$ element.

Interestingly enough, the results of comparison brought in Figure 8 show even better correspondence between the analytical prediction and numerical simulation than the one found in the case of non-linear elastic foundation applied on the chain (cf. Figures 6, 7).

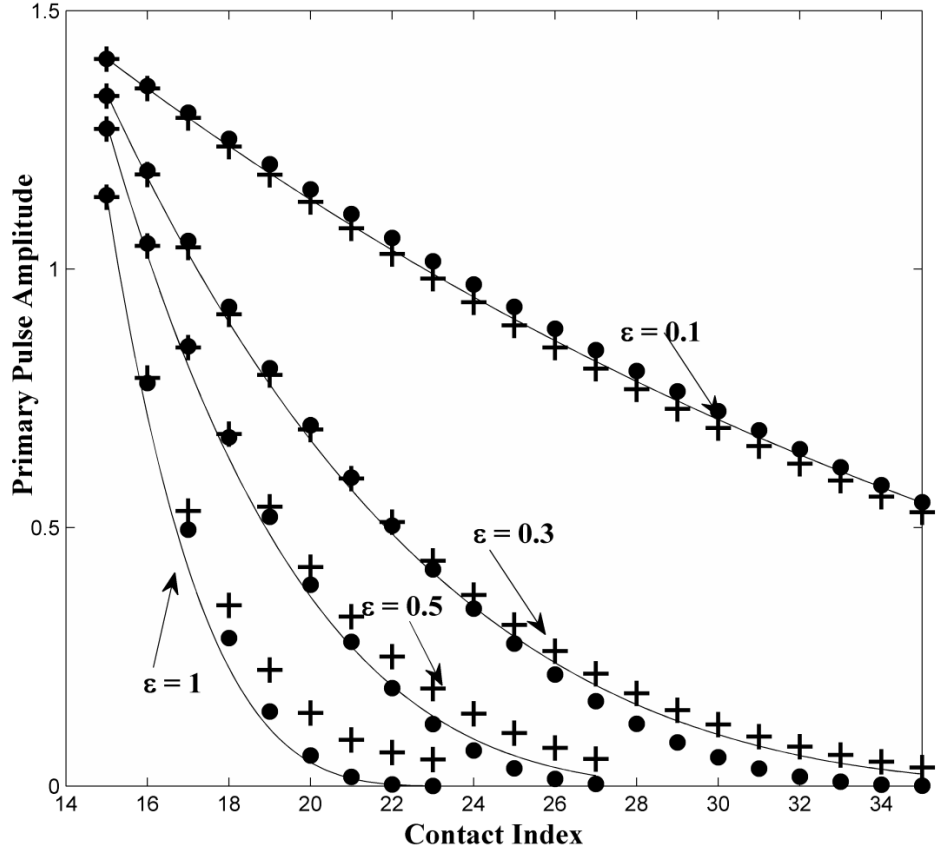

Figure 8 Comparison of analytical prediction and numerical simulation for the various values of $\varepsilon$ parameter ($\varepsilon = 0.1, 0.3, 0.5, 1$), System parameter: $\mu = 1$. Amplitudes of the primary pulse recorded on the contacts of the chain from the direct numerical simulation are denoted with (+) signs, those predicted by the map (35) are denoted with the bolded dots (.) and finally the amplitude evolution predicted by the long wave approximation (36) is denoted with the thin solid line (-). Note that the set of amplitudes $\{A_i\}$ obtained from the map (35) is re-scaled with respect to (32) to provide a fair comparison with the peak of actual response.

4.3 Formation of Stationary Primary Shock Waves in the Perturbed Granular Chains

All the responses concerning the primary pulse propagation through the perturbed granular chain discussed so far are characterized by the monotonous energy radiation from the localized wave front (primary pulse) down to the far field resulting in the constant reduction in the amplitude of the primary pulse. However, as it will be shown in the present section perturbation applied on the granular lattice can be of a somewhat different type leading to the formation of the primary pulses propagating with constant amplitudes and exhibiting solitary like behavior. Basically formation of stationary shocks in the perturbed granular lattices can be easily realized if the perturbation applied on the chain contains the terms leading to the two opposing effects, namely dissipative term and the term acting as an energy source. Therefore, if a perturbation applied on a chain fits to this type, then there is a high possibility for the formation of the aforementioned stationary shocks carrying the nature of solitary waves. As an example for the dynamical system falling under this category one may think of a chain of Van-Der-Pol oscillators coupled through a strong Hertzian contact law. The dynamics of the primary pulse propagating in this system will be addressed herein.

The primary aim of the present sub-section is to demonstrate the potential of the developed methodology in predicting the formation and annihilation of stationary pulses propagating through the perturbed granular chain vs. the variation of the system parameters. For that matter we consider a chain of VDP oscillators coupled via the uniform Hertzian contact law,

$$\ddot{x}_i = \{x_{i-1} - x_i\}^n \theta(x_{i-1} - x_i) - \{x_i - x_{i+1}\}^n \theta(x_i - x_{i+1}) - \varepsilon \left( \dot{x}_i - \mu \left(1 - x_i^2\right) \dot{x}_i \right) \quad (37)$$

Before proceeding with the analysis we plot the response of (37) (obtained from the direct numerical simulations) (Figure 9a,b) illustrating the formation of stationary shock for the two different intensities of the initial impulse applied on the chain. Again we note that numerical procedure applied on System (37) was performed similarly to the ones described in the previous sub-sections. Thus, initial impulse is applied on the left part of the chain where the first 16 elements of the chain are kept unperturbed and the perturbation is only activated starting from the $16^{th}$ element of the chain. We note that the response demonstrated in Figure 9 shows solely the primary edge of the pulse where the residual fluctuations exhibited by the contacts are chopped.

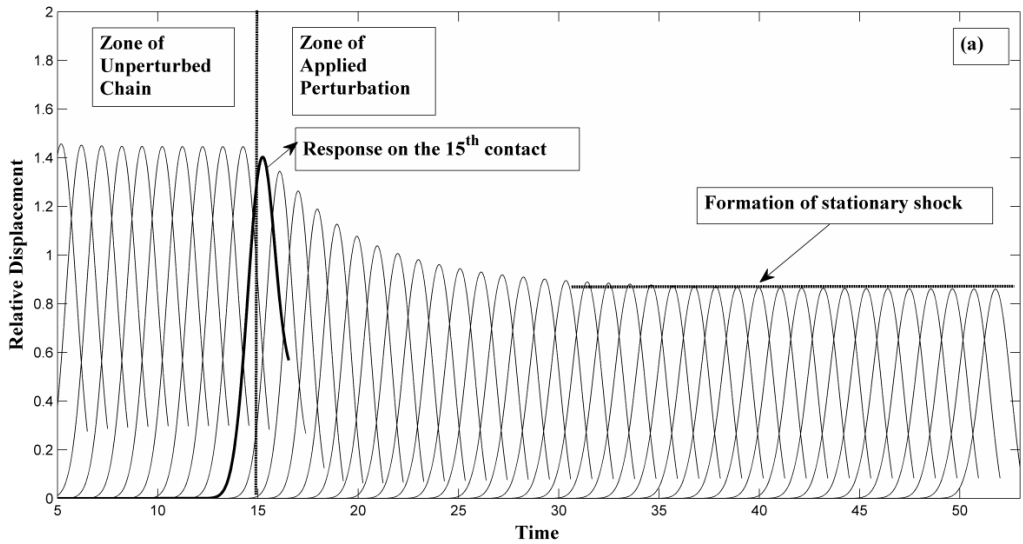

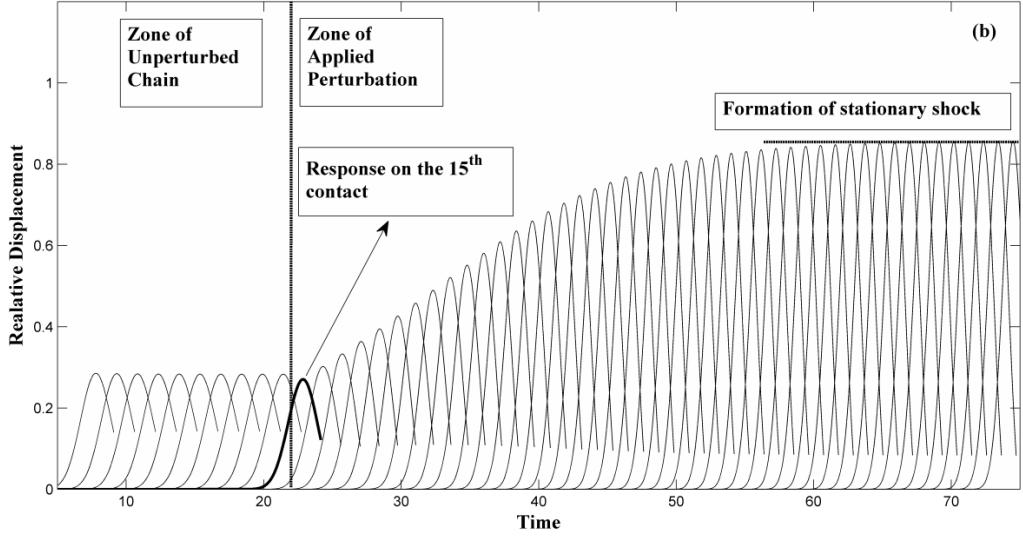

Figure 9 Time series of the response recorded on each one of the contacts for both perturbed and unperturbed part of the granular chains. Only primary pulse of the response recorded on each one of the contacts is shown. System parameter: $\mu = 3, \varepsilon = 0.1$. (a) Higher initial impulse: $\dot{x}_1(0) = 2.3$, (b) Lower initial impulse: $\dot{x}_1(0) = 0.3$

As it is evident from the results of Figure 9 there is a formation of stationary shock propagating with the constant amplitude highly resembling in its nature Nesterenko solitary waves. It is also clear from the response illustrated on Fig. 10 that this stationary shock acts as an attractor for the unperturbed solitary waves impinging on the perturbed part of the chain with different amplitudes.

Note that the parameters of stiffness and nonlinear damping of the VDP oscillator are assumed to be small in comparison with the Hertzian interaction law such that VDP oscillator brings a small perturbation to the Hertzian chain. Arguing as before we transfer System (37) into the coordinates of relative displacements,

$$\ddot{\delta}_i \cong \{\delta_{i-1}\}_+^n - 2\{\delta_i\}_+^n + \{\delta_{i+1}\}_+^n - \varepsilon\left\{\dot{\delta}_i - \mu\dot{\delta}_i + \mu\left((\delta_i + \delta_{i+1})^2(\dot{\delta}_i + \dot{\delta}_{i+1}) - \delta_{i+1}^2\dot{\delta}_{i+1}\right)\right\} \qquad (38)$$

The corresponding map reads,

$$\begin{aligned}
0 &= f_1 A_{i-1}^{5/4} - 2 f_2 A_i^{5/4} + f_3 A_{i+1}^{5/4} + \varepsilon\left\{g_{1000}^{(1)} A_i^{3/4} - \mu g_{0010}^{(1)} A_i + \mu A_i^3 R_1\right\} \\
0 &= (f_5 - f_4) A_i^{5/4} - 2 f_6 A_{i+1}^{5/4} + \varepsilon\left\{g_{10}^{(2)} A_{i+1}^{3/4} - \mu g_{01}^{(2)} A_{i+1} + \mu A_{i+1}^3 R_2\right\}
\end{aligned} \qquad (39)$$

where

$$R_1 = \int_{-\infty}^{0} \left\{ \left(\tilde{S}(t) + \tilde{S}(t-1)\right)^2 \left(\tilde{S}(t)' + \tilde{S}(t-1)'\right) - \left(\tilde{S}(t-1)\right)^2 \left(\tilde{S}(t-1)'\right) \right\} dt, \quad R_2 = \int_{-\infty}^{0} \left\{ \left(\tilde{S}(t-1)\right)^2 \left(\tilde{S}(t-1)'\right) \right\} dt,$$

We note that $R_1, R_2$ were introduced for the sake of brevity; however they can be easily rewritten using the notations of Section 3 ($g_{klmn}^{(1)}, g_{pr}^{(2)}$).

In Figure 10 we illustrate the evolution of the amplitude of the primary pulse as they are recorded on the contacts of the chain subject to the perturbation described in (39). Results of the numerical simulation are compared to the theoretical prediction provided by the map (39) (Figure 10).

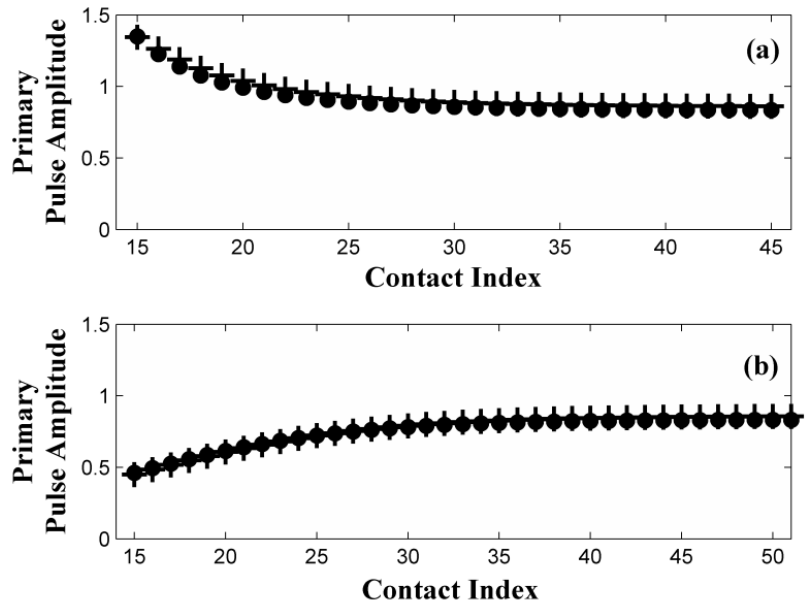

Figure 10 Comparison of analytical prediction and numerical simulation. System parameters: $\varepsilon = 0.1, \mu = 3$ Amplitudes of the primary pulse recorded on the contacts of the chain from the direct numerical simulation are denoted with (+) signs, those predicted by the map (39) are denoted with the bolded dots (.) (a) Higher impulse excitation (b) Lower impulse excitation; Note that the set of amplitudes $\{A_i\}$ obtained from the map (39) is re-scaled with respect to (32) to provide a fair comparison with the peaks of actual response.

As it is clear from the results of Figure 10 there exists a stable attractor corresponding to the propagation of stationary shocks exhibiting solitary like behavior. In fact in the numerical simulation demonstrated in Figure 9 we applied two impulses with different amplitudes and in each one of them the spatial evolution of the amplitude of the pulse gets saturated at a certain point. Thus for the case of a higher amplitude impulse (applied on the perturbed chain) the amplitude of the pulse initially gets reduced till it reaches a certain saturation point (Figures 9a, 10a). As for the case of the lower amplitude impulse the amplitude of the pulse initially climbs up till it reaches the same saturation point (Figures 9b, 10b). These results suggest for the existence of stable attractors (in space of the evolved primary pulses) induced by the perturbation. Simulating the system (39) again (Figure 11); however with different parameter of ($\mu = 1$) we find a monotonous decay of the amplitude of the pulse rather than a stable attractor observed in the previous case for ($\mu = 3$). This may suggest for the absence of any stable attractor.

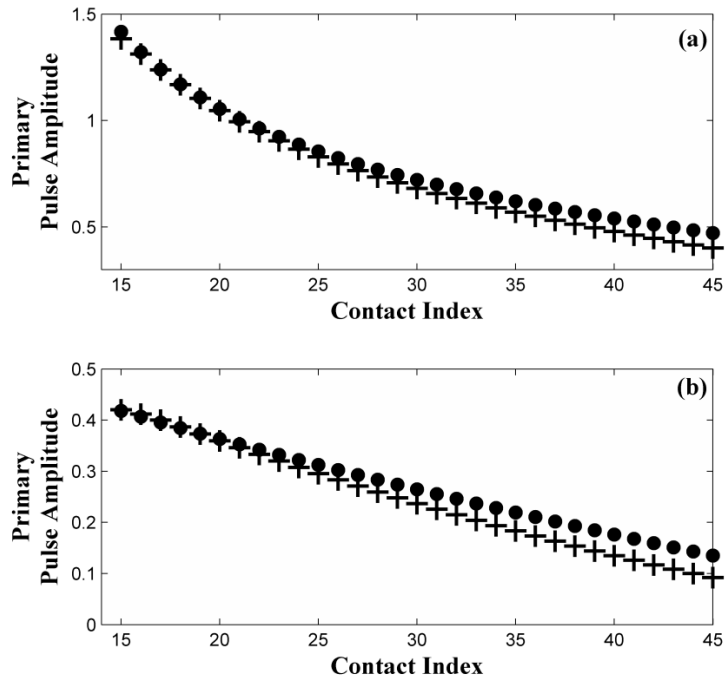

Figure 11 Comparison of analytical prediction and numerical simulation. System parameters: $\varepsilon = 0.1, \mu = 1$ Amplitudes of the primary pulse recorded on the contacts of the chain from the direct numerical simulation are denoted with (+) signs, those predicted by the map (39) are denoted with the bolded dots (.) (a) Higher impulse excitation (b) Lower impulse excitation; Note that the set of amplitudes $\{A_i\}$ obtained from the map (39) is re-scaled with respect to (32) to provide a fair comparison with the peak of the actual response.

Interestingly enough the evolution of the amplitude for both the cases (i.e. existence and absence of the attractors) is perfectly predicted by the map (39) and agrees with the results of numerical simulation (Figures 10, 11). Therefore, the map can be used not only as a tool for the depiction of the evolution of propagating primary pulses but also for the stability analysis of the stationary shocks. To this end let us first find the variation of the amplitude of the attractor with respect to the system parameter $\mu$. In fact the problem of finding the stationary solutions is much simpler than the one dealing with the evolution of the amplitude of the response. To make this statement somewhat more clear let us take several steps back to the construction of the analytical procedure presented in Section 3. Thus, in the first stage of the analytical approach we derived the generalized expression for the map (described by Eq. (19)) directly related to the response on the i$^{th}$ contact. This map assumed the $A_{i-1}$ as known from the previous iteration when $A_i$, $A_{i+1}$ were declared as the two unknowns at the i$^{th}$ step of the map to be found in the successive iterations. Therefore, in order to find $A_i$ at the i$^{th}$ step we required additional (auxiliary) equation to be formulated. This auxiliary equation has been naturally derived from the equation of motion of the (i+1)$^{th}$ contact and for the same time interval (Eq. 22). However, when looking for the stationary regimes (where the amplitude of the primary shock remains constant along the chain for each contact) this additional (auxiliary) equation is not required. Thus, assuming

$$A_{i-1} = A_i = A_{i+1} = A_{i0} \tag{40}$$

right in the first of the equations of (39), immediately yields the following algebraic relation depicting the attractors,

$$g_{1000}^{(1)} A_{i0}^{3/4} - \mu g_{0010}^{(1)} A_{i0} + \mu R_1 A_{i0}^3 = 0 \tag{41}$$

where $A_{i0}$ denotes the amplitude of the attractor. We note that the first three terms of the map can be removed as they correspond to the unperturbed case supporting solitary waves obviously having the same amplitude and thus satisfy, $0 = f_1 A_{i0}^{5/4} - 2 f_2 A_{i0}^{5/4} + f_3 A_{i0}^{5/4}$

In Figure (12) we draw the bifurcation diagram of the stationary shocks with respect to the system parameter $\mu$ described by (41).

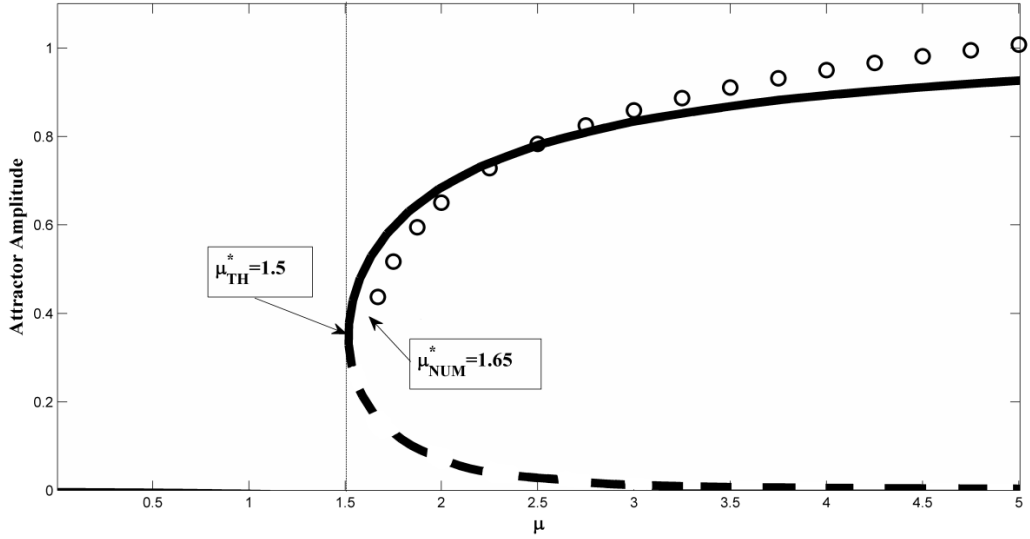

Figure 12 Variation of the amplitude of the attractor with respect to the system parameter $\mu$. System parameters: $\varepsilon = 0.1$. Stable branch of the theoretically predicted response from (41) is denoted with the bold solid line, unstable branch of the response is denoted with the dashed solid line. Amplitudes of the stationary shocks predicted by numerical simulations are denoted with the circles; Note that the amplitude variation depicted by (41) is re-scaled with respect to (32) to provide a fair comparison with the peaks of actual response.

As it is clear from the diagram illustrated on Figure 12 there is a typical saddle node bifurcation overgone by the attractors of the stationary primary pulses leading to the annihilation of the nontrivial solutions. Thus, the proposed methodology of maps predicts the behavior of the primary shock as it propagates through the perturbed granular media. Therefore, if the system parameter $\mu$ is situated below the threshold $\mu^*_{TH} = 1.5$ predicted by the bifurcation diagram (Figure 12), then one would expect for the monotonous decay exhibited by the primary shock (Figure 11). However, in case of the system parameter $\mu$ being selected above the threshold ($\mu > \mu_{TH}$) there is a formation of stationary shocks (Figures 9, 10). It is important to emphasize that the theoretically predicted threshold value $\mu^*_{TH} = 1.5$ is in a close correspondence with that obtained from the numerical simulation $\mu^*_{NUM} \cong 1.65$. Moreover, the results of comparison of the numerically found values of the amplitudes of stationary shocks with the ones derived from the analytical prediction (41) suggest for a very good agreement between the analytical and numerical models (Figure 12).

Thus, concluding the results of the present sub-section we would like to stress once again that the analytical procedure developed in the paper can not only describe the evolution of the amplitude of the primary shock, but can also predict the parametric conditions required for the formation of stable stationary shocks as well as estimating their amplitudes and their stability (Figure 12). However the direct stability analysis for the branches of the attractors of stationary shocks is beyond the scope of the current paper and will be published elsewhere.

4.4. Effect of Randomness of the Elastic Foundation on the Primary Pulse Propagation

As a final example for the application of the proposed methodology we consider the system comprising an uncompressed granular chain mounted on a linear elastic foundation with randomly distributed stiffness coefficients. Equations of motion for the system under consideration read,

$$\ddot{x}_i = \{x_{i-1} - x_i\}^n \theta(x_{i-1} - x_i) - \{x_i - x_{i+1}\}^n \theta(x_i - x_{i+1}) - \varepsilon_i x_i \tag{42}$$

Unlike the previously considered cases the perturbation assumed in (42) is highly non-uniform and is randomly distributed along the chain. Therefore, for each parameter of stiffness $\varepsilon_i$ of the linear elastic foundation applied on the i$^{th}$ element of the chain we assume a uniform random distribution $(\varepsilon_i \sim U[0,1])$.

Again, transformations to coordinates of relative displacements yields,

$$\ddot{\delta}_i \cong \{\delta_{i-1}\}_+^n - 2\{\delta_i\}_+^n + \{\delta_{i+1}\}_+^n - \varepsilon_i \{\delta_i\} \tag{43}$$

The corresponding map constructed for (43) reads,

$$0 = f_1 A_{i-1}^{5/4} - 2 f_2 A_i^{5/4} + f_3 A_{i+1}^{5/4} + \varepsilon_i \left( g_{1000}^{(1)} A_i^{3/4} \right)$$
$$0 = (f_5 - f_4) A_i^{5/4} - 2 f_6 A_{i+1}^{5/4} + \varepsilon_i \left( g_{10}^{(2)} A_{i+1}^{3/4} \right) \tag{44}$$

We note that unlike the case of a uniform distribution considered in previous sub-sections, System (44) cannot be homogenized and therefore any regular analytical approximation rather than the iterative map is possible. In Figure 13 we plot the evolution of the amplitude of the primary pulse for the 4 different realizations of the randomly generated vector containing random stiffness coefficients $\{\varepsilon_i\}$. Again we note that numerical procedure applied on System (42) was performed similarly to the ones described in the previous sub-sections. Thus, initial impulse is applied on the left part of the chain where the first 16 elements of the chain are kept unperturbed and the perturbation is only activated starting from the $16^{th}$ element. Results of the numerical simulation are compared to the theoretical predictions provided by the map (44) (Figure 13).

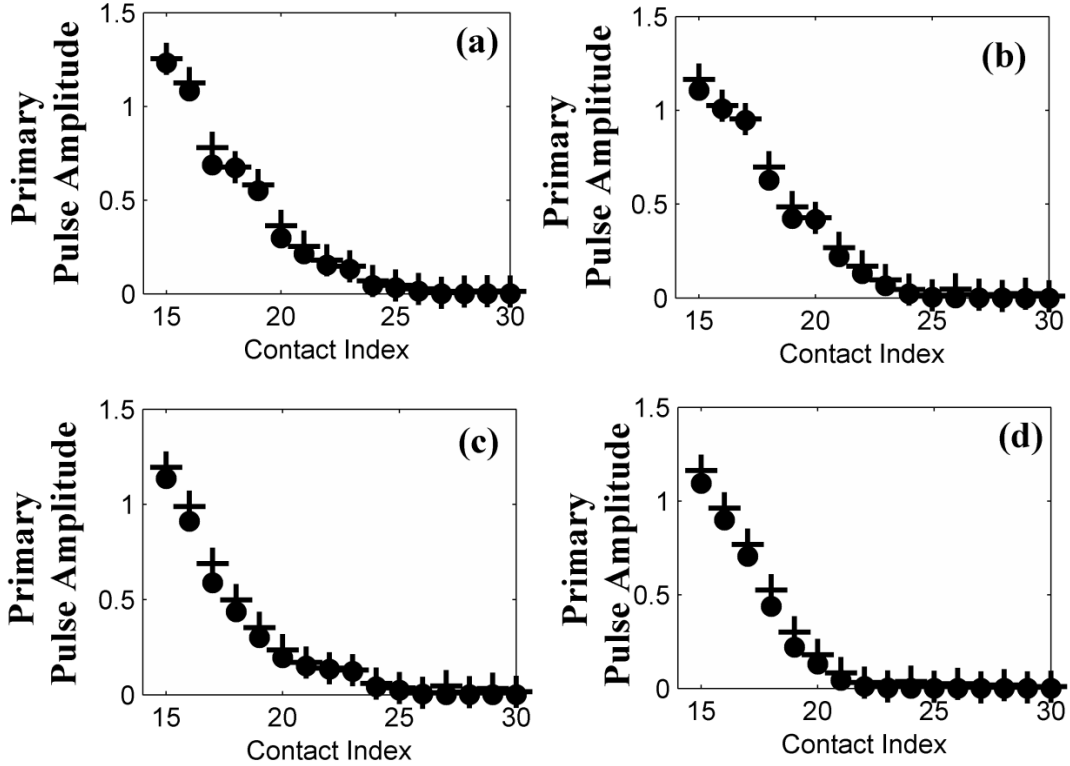

Figure 13 Comparison of analytical prediction and numerical simulation for the random set of $\varepsilon$ parameters ($\varepsilon \sim U[0,1]$). Amplitudes of the primary pulse recorded on the contacts of the chain from the direct numerical simulation are denoted with (+) signs, those predicted by the map (44) are denoted with the bolded dots (.) Note that the set of amplitudes $\{A_i\}$ obtained from the map (44) is re-scaled with respect to (36) to provide a fair comparison with the peak of actual response.

From the results of Figure 13 it is clear that the analytical and numerical models are in a very good agreement. This fact suggests for the robustness of the proposed methodology. Basically the developed procedure carries a potential to be applied on nonlinear lattices subject to various types of random and deterministic perturbations where the estimation of the evolution of a primary pulse under the action of the perturbation is sought. Moreover, authors convinced that the proposed methodology is quite universal and can be farther extended to the general class of nonlinear lattices (e.g. Toda lattices, FPU chains and more) depicting the evolution of the primary pulses.

## 5. Conclusions

Present study concerns the dynamics of primary pulse propagating through the granular chain subject to an external on-site perturbation. Analytical procedure predicting the evolution of the propagating primary pulse is developed for the general form of the on-site perturbation and is based on the construction of the two dimensional maps. The derived maps are further approached by the long limit approximation depicting the global evolution of the pulse through the chain. The developed analytical model is applied on various perturbed granular setups such as, granular chains mounted on the nonlinear elastic foundation of the general type, granular chains perturbed by the dissipative force as well as randomly perturbed chains. Additional interesting finding made in the present study corresponds to the chains subject to a special type of perturbations incorporating the dissipative and energy sourcing terms. It was shown that application of such perturbation may lead to formation of stable stationary shocks acting as attractors for the initially unperturbed, propagating Nesterenko solitary waves. Interestingly enough the developed analytical procedure provides a rather close estimations for the amplitudes of these stationary shocks as well as predicts zones of their stability. As it was shown in the paper for the case of the chain of Van-Der-Pol oscillators coupled via a pure Hertzian contact law, stationary shocks are created through a saddle node bifurcation undergone by a stable and non-stable attractors corresponding to the stationary shocks. In general we must say that the provided analytical model have demonstrated a very good correspondence to the results obtained from the direct numerical simulations and this for all the applications considered in the paper. Authors believe that the developed analytical procedure can be further adjusted to apply for the more complicated types of nonlinear lattices and is of broad applicability for any physical problem seeking for the estimation of the modulation of localized nonlinear waves (e.g. solitons, breathers, localized wave-packets, etc.) in the perturbed, 1D discrete media.

**Acknowledgements** The generous financial support of the Technion – Israel Institute of Technology is Gratefully Acknowledged